\documentclass{article}
\usepackage[T2A]{fontenc}
\usepackage{amsmath}
\usepackage{amssymb}
\usepackage{color}

\usepackage{hyphenat}
\usepackage[english]{babel}

\newcounter{dummy} \numberwithin{dummy}{section}
\newtheorem{theorem}[dummy]{Theorem}
\newtheorem{lemma}[dummy]{Lemma}

\newtheorem{proposition}[dummy]{Proposition}
\newtheorem{remark}[dummy]{Remark}
\def\eps{ { \varepsilon } }
\def\phi{{\varphi}}
\DeclareMathOperator{\sgn}{sgn}

\newcommand{\dco}{\mathcal{M}}
\newcommand\at[2]{\left.#1\right|_{#2}}
\newcommand*{\QEDA}{\hfill\ensuremath{\blacksquare}}
\numberwithin{equation}{section}
\usepackage
[
        a4paper,% other options: a3paper, a5paper, etc
        left=3cm,
        right=3cm,
        top=3cm,
        bottom=4cm,
]
{geometry}
\usepackage{authblk}
\author[1,2]{A. Elokhin}
\affil[1]{National Research University "Higher School of Economics"}
\affil[2]{Steklov Mathematical Institute of Russian Academy of Sciences}
\title{Asymptotics for a class of singular integrals of quotients with highly degenerate denominators}
\begin{document}
\maketitle
\begin{abstract}
In rigorous study of stochastic models for the wave turbulence theory and R. Peierls's kinetic theory for the thermal conductivity in solids, analysis of integrals of the form $\int_{\dco}\frac{F\omega_\dco}{\Omega^2 + \nu^2\Gamma^2}$ and $\int_{\dco}\frac{F\cos(\nu^{-1}\Omega)\omega_\dco}{\Omega^2 + \nu^2\Gamma^2}$ plays a crucial role, where $\nu>0$ is a small parameter, $\dco$ is a closed Riemannian manifold with volume form $\omega_\dco$, and the functions $\Gamma > 0$, $F$, $\Omega$ are sufficiently smooth. We investigate the asymptotic behavior of the integrals in the limit $\nu\rightarrow 0$. This work continues studies [Kuksin' 17, Dymov' 23], in which the authors considered similar integrals for the case $\dco=\mathbb{R}^d$ when the function $\Omega$ is Morse. We significantly weaken the latter assumption, which played an important role in the aforementioned works. This makes the obtained results applicable to the problem of rigorous justification of R. Peierls's kinetic theory.
\end{abstract}
\section{Introduction}
\subsection{Setting and results}
Let $\dco$ be a sufficiently smooth closed Riemannian manifold of dimension $d \geq 4$, equipped with a Riemannian metric $g$. Consider the integrals
\begin{equation}\label{int_study}
\int_{\dco}\frac{F\omega_\dco}{\Omega^2 + \nu^2\Gamma_A^2},\qquad \int_{\dco}\frac{F\cos(\rho\nu^{-1}\Omega)\omega_\dco}{\Omega^2 + \nu^2\Gamma_B^2},
\end{equation}
where $\omega_\dco$ is a volume form on the manifold $\dco$, real-valued functions $\Omega$, $\Gamma_A$ and $F$ on $\dco$ are assumed to be sufficiently smooth, the function $\Gamma_A$ is strictly positive, and constants $\Gamma_B>0$, $\rho \in \mathbb{R}$. We are interested in the asymptotic behavior of integrals \eqref{int_study} in the limit $\nu \rightarrow 0$. Analysis of these integrals plays an important role in two famous problems of mathematical physics: in the problem of rigorous justification of the wave turbulence theory and of R. Peierls's kinetic theory for the thermal conductivity in crystals. The behavior of integrals \eqref{int_study} was previously studied in \cite{SK}, \cite{SK2}, \cite{integrals} for the case $\dco = \mathbb{R}^d$. In \cite{SK} and \cite{SK2} the first and second integrals from \eqref{int_study} were investigated, respectively, in the important particular case when the function $\Omega$ is a non-degenerate quadratic form while work \cite{integrals} deals with general Morse functions $\Omega$. The assumption that critical points of $\Omega$ are non-degenerate plays a significant role in these works. However, for a number of applications, this requirement turns out to be too restrictive. In particular, the result from \cite{integrals} is not applicable to the aforementioned problem of rigorous justification of Peierls's theory; see Section 1.2 for a more detailed discussion.

Let
$$\Sigma = \Omega^{-1}(\{0\}),$$
and assume that $\Sigma \neq \emptyset$ (otherwise analysis of integrals \eqref{int_study} as $\nu\rightarrow 0$ is trivial). Let
\begin{equation}\label{psi_set}
\Psi := \{x\in\Sigma: d\Omega(x) = 0\}
\end{equation}
be a set of critical points of the function $\Omega$ that belong to $\Sigma$. We will assume that for any point $a \in \Psi$, there exists a neighborhood $U(a) \subset \dco$ such that for any point $b \in U(a)$

\begin{equation}\label{grad_cond_manifold}
||d\Omega(b)||_g \geq C \text{dist}(a,b)^q,\qquad q\geq 1,
\end{equation}
where $||v||_g = \sqrt{\sum_{i,j}{g^{ij}(b)v_i v_j}}$, $v \in T^*_b\dco$, is the norm induced by the metric $g$, and $g^{ij}$ are components of the inverse metric tensor; $\text{dist}(a,b)$ denotes the distance\footnote{Computed as the length of the shortest geodesic connecting the points $a$ and $b$.} between points $a, b \in \dco$, while the parameter $q \in [1, \infty)$ and the constant $C$ are independent of the choice of points $a,b$. Note that, due to \eqref{grad_cond_manifold} and the compactness of $\dco$, the critical points of $\Omega$ on $\Sigma$ are isolated, and their number is finite. Also note that for $q=1$, assumption \eqref{grad_cond_manifold} is equivalent to non-degeneracy of all critical points belonging to $\Sigma$. However, for $q>1$ this is not the case.

Dividing the numerator and denominator in \eqref{int_study} by $\Gamma^2_{A,B}$, we obtain integrals of the same form with $\Gamma_{A,B} \equiv 1$ and modified functions $F$, $\Omega$ and constant $\rho$. Moreover, after this reduction, the integral $I^A_\nu$ becomes a special case of $I^B_\nu$ with $\rho=0$. Thus, it suffices to study the integral
\begin{equation}\label{int_study_div}
I_\nu := \int_{\dco}\frac{F\cos(\rho\nu^{-1}\Omega)\omega_\dco}{\Omega^2 + \nu^2}.
\end{equation}
By the implicit function theorem, the set $\Sigma_0 = \Sigma \setminus \Psi$ is a differentiable manifold of dimension $d-1$. Denoting by $\omega_{\Sigma_0}$ the volume form on $\Sigma_0$ induced from $(\dco, \omega_\dco)$, we consider the integral
\begin{equation}\label{int_asymp}
I_0 := \pi \int_{\Sigma_0} \frac{F \omega_{\Sigma_0}}{||d\Omega||_{g}}.
\end{equation}
Let $r > 0$, $a, b \in \mathbb{R}$. We define the function
\begin{equation*}
\chi_{a,b}(r) =
\begin{cases}
1, & a \neq b, \\
|\ln(r)|, & a = b.
\end{cases}
\end{equation*}
We now state the main result of this work.
\begin{theorem}\label{th_main}
	Let $\dco$ be a $C^2$-smooth manifold of dimension
    $$d\geq 4q,\quad q\in[1,\infty),$$
    equipped with a $C^2$-smooth Riemannian metric $g$. Assume that for every point of the set $\Psi$, defined in \eqref{psi_set}, there exists a neighborhood in which assumption \eqref{grad_cond_manifold} is satisfied. Furthermore, let the function $F$ be $C^2$-smooth and $\Omega$ be $C^5$-smooth. Then,
    \begin{equation}\label{statement}
    |I_\nu - \nu^{-1} e^{-|\rho|} I_0| \leq C \chi_{d,4q}(\nu),
    \end{equation}
    where the constant $C$ does not depend on $0 < \nu \leq 1/2$.
\end{theorem}
Straightforward analysis of the proof of Theorem \ref{th_main} implies the following result, used in \cite{ours}.
\begin{remark}
Let $\mathcal K$ be a compact set. Suppose that the functions $\Omega$, $F$ in \eqref{int_study_div} depend on a parameter $k\in \mathcal K$, are continuous as the functions of $(k,x)$ and that the assumptions of Theorem \ref{th_main} are satisfied. In particular, assumption \eqref{grad_cond_manifold} holds with constants $C$ and $q$ independent from $k\in \mathcal K$. Then, the constant $C$ from \eqref{statement} can be chosen independently of $k$ as well.
\end{remark}

\subsection{Motivation}

The study of the asymptotics of integrals \eqref{int_study}, where $\dco$ is a closed Riemannian manifold or $\mathbb{R}^n$, plays a central role in the rigorous analysis of two famous kinetic theories for dispersive nonlinear wave systems in stochastic settings --- R. Peierls's kinetic theory of the thermal conductivity in solids and the wave turbulence theory. The former was developed in 1929 \cite{peierls}, while the latter was established in the 1960s and since then it has been actively developed in the physics literature, see \cite{nazarenko, zakharov}. However, the problem of providing a mathematical foundation for the core principles of the wave turbulence theory has only begun to advance in the last decade, see e.g. \cite{deng_hani, wt_text1, spohn} and references therein, while the rigorous justification of the theory of Peierls remains an open problem. 

The both theories study low-amplitude solutions for dispersive Hamiltonian systems of a large number of nonlinear interacting waves. Central objects in the both theories are nonlinear kinetic equations describing the behavior at mesoscopic scale of certain physical observables computed for solutions of the original systems.

Components of the kinetic kernels $K$, that define the principal terms of the kinetic equation, are given by integrals of the form \eqref{int_asymp},
where the functions $F = F_k$ and $\Omega = \Omega_k$ depend on a parameter $k\in\dco$. The function $\Omega_k$, written in the appropriate local coordinates, takes the form
\begin{equation}\label{delta_omega}
\begin{split}
& \Omega_k(y_1,...,y_{N-2}) := \sum_{i=1}^p\lambda(y_i) - \sum_{i=p+1}^{N-1}\lambda(y_i) - \lambda(k),
\end{split}
\end{equation}
where $y_{N-1}$ is expressed from the equation
$$\sum_{i=1}^p y_i = \sum_{i=p+1}^{N-1} y_i + k,$$
the number $p$ satisfies $0\leq p\leq N-1$, $N\geq 3$ is the number of waves participating in each interaction, $\lambda$ is the dispersion relation of the original wave system and $y = (y_1,...,y_{N-1})$, $y_i\in\mathbb{R}^n$, $n\geq 2$; see \cite[Sections 6.9, 6.11]{nazarenko} for more details.

One approach to studying such Hamiltonian systems involves adding weak random noise and viscosity, governed by a small parameter $\nu$, to the original equations of motion. In this setting, nonlinearities in the equations that describe the dynamics of the physical observables at microscopic scale are approximated by integrals \eqref{int_study}. Then, the kinetic kernel $K$ in the kinetic equation is obtained by passing to the limit $\nu\rightarrow 0$. The objective of the present work is to justify this limit.

The most well-studied system in the context of the wave turbulence theory is the cubic nonlinear Schrödinger equation, see e.g. \cite{deng_hani, wt_text1, spohn} and references therein. In this case, the dispersion relation takes the form
$$\lambda(y) = |y|^2,\quad y\in\mathbb{R}^n,$$
so the function $\Omega_k$ is a non-degenerate quadratic form. The corresponding asymptotic analysis of integrals \eqref{int_study} was carried out in \cite{SK} and \cite{SK2}, utilizing the explicit form of the function $\Omega_k$, and the obtained result was applied in \cite{wt_text1}.

A more complex form of the dispersion relation arises, for example, when considering the Petviashvili or Charney–Hasegawa–Mima equations. The result of work \cite{integrals} applies, in particular, to these cases. However, it crucially uses that the function $\Omega_k$ does not have degenerate critical points on the set $\Sigma$. The latter requirement is not satisfied in the Peierls model \cite{spohn, peierls}, which studies energy transport in a lattice of oscillators with weakly nonlinear nearest-neighbor interaction. The "classical" dispersion relation for the Peierls's model has the form
\begin{equation}\label{omega_peierls}
\lambda(y) = \sqrt{\sum_{i=1}^n\sin^2{\frac{y^i}2}},\quad y\in \mathbb{T}^n,
\end{equation}
where $\mathbb{T}^n$ denotes the $n$-dimensional torus with period $2\pi$. It can be shown that in this case, for some values of the parameter $k$, the function $\Omega_k$, corresponding to dispersion relation \eqref{omega_peierls}, has degenerate critical points. However, a more subtle analysis of the critical points is complicated since their explicit form is not known.

Together with the "classical" Peierls's model, the study of systems with modified dispersion relations and dispersion relations of a general form is of great interest \cite{kupiainen, spohn2}. Assumption \eqref{grad_cond} in this context is natural in difference with the assumption that $\Omega_k$ does not have degenerate critical points. Indeed, one can see that the latter situation in not generic as $\Omega_k$ depends on the parameter $k$. In particular, in Appendix we show that assumption \eqref{grad_cond} with $q=2$ is satisfied for the considered in \cite{kupiainen} dispersion relation 
\begin{equation}\label{omega_kupiainen}
\lambda^{*}(y) = \sum_{i=1}^n\sin^2{\frac{y^i}2} + m,\quad y\in \mathbb{T}^n,\quad m>0,
\end{equation}
and $\Omega_k$ is given by \eqref{delta_omega} with $N=4$, $p=2$. In \cite{ours}, relying on the result of the present paper, we study the Peierls's model assuming that the dispersion relation is such that \eqref{grad_cond} holds. 

\subsection{Preliminaries}
Without loss of generality we assume that set \eqref{psi_set} contains at most one point, and that the support of the function $F$ from \eqref{int_study_div} lies entirely within a single chart. Indeed, due to the compactness of the manifold $\dco$, there exists a finite open cover $\{U_i\}_{1 \leq i \leq n}$ and a smooth partition of unity $\{f_i\}_{1\leq i \leq n}$, subordinate to this cover \cite[Theorem 13.7]{manifolds}, such that each set $U_i \cap \Psi$ contains at most one point, and each $U_i$ is entirely contained in a single chart. Accordingly, it suffices to justify asymptotics \eqref{statement} for integrals \eqref{int_study_div} in which the function $F$ is replaced by $F f_i$.

Let us choose local coordinates on the chart containing $\text{supp}\; F$:
\begin{equation}\label{loc_coord}
x = (x^1, \ldots, x^d) \in \mathbb{R}^d,
\end{equation}
and denote by $D$ an open subset of $\mathbb{R}^d$ such that $\text{supp}\; F \subset D$. Without loss of generality, we assume that $|x| < 1$ for any $x \in D$, where $|\cdot|$ denotes the Euclidean norm in $\mathbb{R}^d$. It is convenient to choose the set $D$ in such a way that
\begin{equation}\label{D_property}
|x-\Sigma| < |x - \Sigma \cap\partial D|,\quad \forall x\in D.
\end{equation}
To this end, we can choose a sufficiently large open set $D$ such that $\operatorname{supp} F \subset D$ and then remove from $D$ a closed subset of points for which \eqref{D_property} is not satisfied.

We restrict our consideration to the case when $\Sigma_0 \cap D$ is a differentiable $(d-1)$-dimensional manifold, and the only critical point of the function $\Omega$ lying on $\Sigma \cap D$ is $x = 0$. Other possible cases can be obtained by simplifying the arguments presented below. 

In the chosen coordinates, the integral $I_\nu$ takes the form
\begin{equation}\label{int_chart}
I_\nu = \int_{D}\frac{F(x)\cos(\rho\nu^{-1}\Omega(x))\sqrt{\det g(x)}}{\Omega^2(x) + \nu^2}dx,
\end{equation}
where we used the notation $dx := dx^1\wedge ...\wedge dx^d$. Assumption \eqref{grad_cond_manifold} takes the form
\begin{equation}\label{grad_cond}
|\nabla\Omega(x)| \geq C|x|^q, \quad\text{where}\quad \nabla\Omega(x) = (\partial_{x^1}\Omega,...,\partial_{x^d}\Omega),
\end{equation}
and without loss of generality we assume that \eqref{grad_cond} holds for all $x\in D$. Moreover, due to the $C^2$-smoothness of $\Omega$ and the boundedness of $D$, we also have the estimate
\begin{equation}\label{grad_upper}
|\nabla\Omega(x)|\leq C|x|,\qquad \forall x\in D.
\end{equation}
Abusing notation, below we redefine 
\begin{equation}\label{abusing_notation}
\Sigma := \Sigma\cap D,\quad \Sigma_0 := \Sigma\setminus\{0\}.
\end{equation}

The proof of Theorem \ref{th_main} reduces to the analysis of integral \eqref{int_chart} over three subsets of $D$: a neighborhood of the critical point $x=0$, a neighborhood of the manifold $\Sigma_0$, and the intersection of their complements. We will show that the asymptotics comes only from the integral over the neighborhood of $\Sigma_0$, while the integral over the remaining part of $D$ is small. 
 
\section{Analysis of the manifold $\Sigma_0$ and its neighborhood}
In this and the next sections, we will consider the manifolds $\Sigma$ and $\Sigma_0$ (recall \eqref{abusing_notation}) as subsets of the space $\mathbb{R}^d$ with the standard Euclidean structure. We will denote various constants by $C$, $C_1$,~\ldots, which, unless otherwise stated, are independent of $\nu$ but may depend on the parameter $\Theta$ introduced below.

Let us introduce differentiable coordinates $\xi \in \mathbb{R}^{d-1}$ on the manifold $\Sigma_0 \subset \mathbb{R}^d$ and consider the functions $\xi \rightarrow x_\xi \in \Sigma_0$. For convenience, we will sometimes also write $\xi \in \Sigma_0$. Define the normal vector to the manifold $\Sigma_0$ at the point $\xi$
\begin{equation}\label{norm_vector}
N_\xi = \nabla \Omega(x_\xi),\quad \nabla\Omega = (\partial_{x^1}\Omega,...,\partial_{x^d}\Omega).
\end{equation}
Consider the mapping $\pi : \Sigma_0\times\mathbb{R}\rightarrow \mathbb{R}^d$, defined by
\begin{equation}\label{pimap} 
	\pi(\xi,\theta) = x_\xi + \theta N_\xi.
\end{equation}
Consider a neighborhood of the manifold $\Sigma_0$
$$U_\Theta(\Sigma_0) = \{x = \pi(\xi,\theta)\in D:\xi\in\Sigma_0,|\theta|<\Theta\},$$
where $\Theta \in (0, 1]$ is assumed to be sufficiently small but independent of $\nu$. 
\begin{lemma}\label{unique_param}
	For any sufficiently small $\Theta$, the mapping $\pi: \Sigma_0 \times[-\theta,\theta]\rightarrow U_\Theta(\Sigma_0)$ is a bijection. Moreover, the function $\Omega$ admits the representation
	\begin{equation}\label{omega_repr}
		\Omega(x) = \theta|N_\xi|^{2}h_\xi(\theta),\qquad \forall x = \pi(\xi, \theta) \in U_\Theta(\Sigma_0),
	\end{equation}
	where the $C^3$-function $\theta \rightarrow h_\xi(\theta)$ satisfies $h_\xi(0) = 1$, and the following estimates hold uniformly in $\xi$ and $\theta$
	\begin{equation}\label{g_property}
	|h_\xi(\theta) - 1|\leq C\theta,\quad\left|\frac{d^k}{d\theta^k}h_{\xi}(\theta)\right|\leq C,\quad\text{for}\quad k=1,2,3.
	\end{equation}
\end{lemma}
\textit{Proof.} Let us show that for a sufficiently small $\Theta$ with $\xi_i \in \Sigma_0$ and $|\theta_i| < \Theta$, $i = 1, 2$, the equality $\pi(\xi_1, \theta_1) = \pi(\xi_2, \theta_2)$ implies $x_{\xi_1} = x_{\xi_2}$ and $\theta_1 = \theta_2$. From \eqref{pimap} we have
\begin{equation}\label{pimap_diff}
x_{\xi_1} - x_{\xi_2} = \theta_2 N_{\xi_2} - \theta_1 N_{\xi_1}.
\end{equation}
Taking the scalar product of the both sides with $x_{\xi_1} - x_{\xi_2}$ yields
\begin{equation}\label{sm_diff}
|x_{\xi_1} - x_{\xi_2}|^2 = \langle\theta_2 N_{\xi_2}, x_{\xi_1} - x_{\xi_2}\rangle - \langle \theta_1 N_{\xi_1}, x_{\xi_1} - x_{\xi_2}\rangle.
\end{equation}
Let $||\cdot||$ denote the operator norm in $\mathbb{R}^d$. Expanding $\Omega(x_{\xi_2})$ using Taylor's formula around $x_{\xi_1}$ and noting that $\Omega(x_{\xi_1}) = \Omega(x_{\xi_2}) = 0$, we obtain
\begin{equation}\label{phi_change}
|\langle \nabla\Omega(x_{\xi_1}), x_{\xi_1} - x_{\xi_2}\rangle|\leq \frac12\max_{x\in[x_{\xi_1},x_{\xi_2}]}||\text{Hess}(\Omega(x))||\;|x_{\xi_1} - x_{\xi_2}|^2\leq C |x_{\xi_1} - x_{\xi_2}|^2,
\end{equation}
where $[x_{\xi_1}, x_{\xi_2}]$ denotes the line segment connecting points $x_{\xi_1}$ and $x_{\xi_2}$, and the last inequality holds due to the compactness of the manifold $\dco$. The constant $C$ in the last inequality is independent of $\Theta$. Writing a similar estimate for the point $x_{\xi_2}$ and taking into account \eqref{norm_vector}, we get
\begin{equation*}
|\langle N_{\xi_i}, x_{\xi_1} - x_{\xi_2}\rangle|\leq C|x_{\xi_1} - x_{\xi_2}|^2,\quad i=1,2.
\end{equation*}
Then from \eqref{sm_diff} it follows that

\begin{equation}\label{square_estimate}
	|x_{\xi_2} - x_{\xi_1}|^2\leq C\Theta|x_{\xi_1} - x_{\xi_2}|^2.
\end{equation}
Choosing $\Theta < C^{-1}$, we obtain $x_{\xi_1} = x_{\xi_2}$, and then from \eqref{pimap_diff} it follows that $\theta_1 = \theta_2$, as required.

To prove relations \eqref{omega_repr}, \eqref{g_property} we will fix $\xi$ and apply Taylor's formula to the function $\theta \rightarrow \Omega_\xi(\theta) = \Omega(\pi(\xi, \theta))$ at the point $\theta = 0$. We have
$$\Omega'_\xi(0) = \at{\frac{d}{d\theta}}{\theta=0}\Omega(x_\xi + \theta N_{\xi}) = \langle\nabla\Omega(x_\xi),N_\xi\rangle= |N_\xi|^{2},\qquad\Omega''_\xi(\theta)=\langle\text{Hess}(\Omega(x_\xi + \theta N_\xi)N_\xi),N_\xi\rangle.$$ 
Then, since $\Omega_\xi(0) = 0$, we obtain
\begin{equation*}
	\begin{split}
		&\Omega_\xi(\theta) = \theta|N_\xi|^{2} + \int_0^\theta\langle\text{Hess}(\Omega(x_\xi +  t N_\xi))N_\xi),N_\xi\rangle(\theta - t)dt\\
		&=\theta|N_\xi|^{2}\left[1 + \theta\int_0^1\langle\text{Hess}(\Omega(x_\xi +  s\theta N_\xi))n_\xi),n_\xi\rangle(1-s)ds\right],
	\end{split}
\end{equation*}
where $n_\xi := {N_\xi}/{|N_\xi|}$. Thus, we get \eqref{omega_repr} with
\begin{equation}\label{h_def}
h_\xi(\theta) = 1 + \theta\int_0^1\langle\text{Hess}(\Omega(x_\xi +  s\theta N_\xi))n_\xi),n_\xi\rangle(1-s)ds.
\end{equation}
Since the function $\Omega$ is $C^5$-smooth, the function $h_\xi(\cdot)$ is $C^3$-smooth. Estimates \eqref{g_property} follow from the uniform boundedness of the integrand and its first three derivatives with respect to $\theta$, uniformly in $\xi$, $\theta$ and $s$.

\QEDA

\begin{proposition}\label{prop_volume_element}
    The volume element $dx$ on the set $U_\Theta(\Sigma_0)$ is written in coordinates $(\xi, \theta)$ as
    \begin{equation}\label{volume_element}
    dx = |N_\xi| \mu_\xi(\theta) d\theta m(d\xi),
    \end{equation}
    where $m(d\xi)$ denotes the volume element on $\Sigma_0$ induced by the standard Euclidean structure of $\mathbb{R}^d$. The function $\mu_\xi$ satisfies
    \begin{equation}\label{mu_ineq}
    \mu_\xi(\theta) \geq C^{-1} > 0, \qquad \mu_\xi(0) = 1, \qquad \left|\frac{d^k}{d\theta^k}\mu_{\xi}(\theta)\right| \leq C
    \end{equation}
    for $0 \leq k \leq 2$, uniformly in $\xi \in \Sigma_0$ and $|\theta| < \Theta$.
\end{proposition}
\textit{Proof.} Consider a projection $\Pi$, mapping $x = \pi(\xi, \theta)$ to $x_\xi = \pi(\xi, 0)$,
\begin{equation}\label{proj}
\Pi: U_\Theta(\Sigma_0) \rightarrow \Sigma_0, \quad \Pi(x) = x_\xi.
\end{equation}
The kernel of the differential $d\Pi(x)$ at the point $x = \pi(\xi, \theta)$ is given by $\text{span}(N_\xi)$. Denote by $J_{\Pi}(x)$ the Jacobian of the mapping $d\Pi(x)$ restricted to the orthogonal complement $(\text{span}(N_\xi))^\perp$, and apply the coarea formula \cite[Theorem 13.4.2]{coarea} for an arbitrary Lebesgue measurable set $A \subset U_\Theta(\Sigma_0)$. We obtain
$$\int_A J_{\Pi}(x)dx = \int_{\Sigma_0}l(A\cap\Pi^{-1}(x_\xi))m(d\xi),$$
where $l(\cdot)$ denotes the Lebesgue measure on the set $\Pi^{-1}(x_\xi) = \{x = \pi(\xi, \theta) : |\theta| < \Theta\}$.
We have
$$l(A\cap\Pi^{-1}(x_\xi)) = |N_\xi|\;l(\{\theta:\pi(\xi,\theta)\in A\}).$$
Since the set $A$ is arbitrary, we obtain $J_{\Pi}(x)dx = |N_\xi|d\theta m(d\xi)$. Hence, the formula \eqref{volume_element} is proven with the density function $\mu_\xi$ given by
$$\mu_\xi(\theta) = \frac{1}{J_\Pi(\xi,\theta)},\quad \text{where}\quad J_\Pi(\xi,\theta) := J_{\Pi}(\pi(\xi,\theta)).$$

It remains to prove inequalities \eqref{mu_ineq}, for which we need to derive an explicit form of the Jacobian $J_\Pi(\xi,\theta)$. We fix a point $x =\pi(\xi,\theta)$ and a vector $v\perp N_\xi$, and choose a small $t\in\mathbb{R}$ such that $x + vt\in U_\Theta(\Sigma_0)$. Then, $\pi(\xi,\theta) + vt = \pi(\xi(t),\theta(t))$, where the functions $\xi(t)$ and $\theta(t)$ are uniquely determined for small $t$ by the vector $v$ and the point $x$, satisfying $\xi(0) = \xi$ and $\theta(0) = \theta$. Using \eqref{pimap} and \eqref{proj}, we write
$$\Pi(x) + \theta N_\xi + vt = \Pi(x+vt) + \theta(t)N_{\xi(t)}.$$
Differentiating this relation with respect to $t$ at $t=0$, we obtain
\begin{equation}\label{map_diff}
v = d\Pi(x)v + \frac{d\theta}{dt}\bigg|_{t=0}N_\xi + \theta\frac{dN_{\xi(t)}}{dt}\bigg|_{t=0}.
\end{equation}
By \eqref{norm_vector}, the derivative in the last term is given by
$$\frac{dN_{\xi(t)}}{dt}\bigg|_{t=0} = \frac{d}{dt}\bigg|_{t=0}\nabla\Omega(\Pi(x + vt)) = \text{Hess}(\Omega(x_\xi))(d\Pi(x)v).$$
Next, let $\text{Pr}: \mathbb{R}^d\rightarrow \mathbb{R}^{d-1}$ be the projection operator onto the orthogonal complement of $N_\xi$. Since the vectors $v$ and $d\Pi(x)v$ are orthogonal to the vector $N_\xi$, applying $\text{Pr}$ to \eqref{map_diff} yields
$$v = (\text{Id}_{d-1} + \theta\;\text{Pr}\circ\text{Hess}(\Omega(x_\xi)))d\Pi(x)v,$$
where the vectors $v$ and $d\Pi(x)v$ are viewed as vectors in the $(d-1)$-dimensional space, and $\text{Id}_{d-1}$ is the identity operator. Since the operator norm of the Hessian matrix is bounded uniformly in $\xi$, the operator
$$Q(\xi,\theta) := \text{Id}_{d-1} + \theta\;\text{Pr}\circ\text{Hess}(\Omega(x_\xi))$$
is invertible for any $\xi$ and $|\theta| < \Theta$, once $\Theta$ is sufficiently small. Therefore,
$$d\Pi(x) = (Q(\xi,\theta))^{-1},\quad J_\Pi(\xi,\theta) = \frac{1}{\det Q(\xi,\theta)},$$
which implies $\mu_\xi(\theta) = \det Q(\xi,\theta)$. Thus, the function $\mu_\xi(\theta)$ is $C^2$-smooth and its derivatives are bounded uniformly in $\xi$. It is separated from zero due to the uniform invertibility of the matrix $Q$, and for any $\xi$ we have $\mu_\xi(0) = \det{\text{Id}_{d-1}} = 1$, which concludes the proof.

\QEDA

Let $B_\delta\subset\mathbb{R}^d$ denote the ball of radius $\delta$ centered at the origin. Recall that the parameter $q$ is defined in \eqref{grad_cond}.
\begin{lemma}\label{lemm_int_over_sigma}
Let $p \leq d - q$. Then for any $\delta\in(0,1/2]$,
\begin{equation}\label{int_lemm}
\int_{\Sigma_0\setminus B_\delta} |x_\xi|^{-p}m(d\xi)\leq C\chi_{d, p+q}(\delta).
\end{equation}
Moreover, for $p < d - q$
\begin{equation}\label{int_lemm_2}
\int_{\Sigma_0\cap B_\delta} |x_\xi|^{-p}m(d\xi)\leq C_1\delta^{d - q - p}.
\end{equation}
\end{lemma}
\textit{Proof.} We will use the estimate for $x = \pi(\xi,\theta)\in U_\Theta(\Sigma_0)$
\begin{equation}\label{supp_est}
C^{-1}|x_\xi|\leq |x| \leq C |x_\xi|,
\end{equation}
which follows from estimate \eqref{grad_upper} and the triangle inequality. Consider the trivial identity
$$m(d\xi) = \frac{1}{2\Theta}\int_{-\Theta}^\Theta\frac{|N_\xi|\mu_\xi(\theta)d\theta m(d\xi)}{|N_\xi|\mu_\xi(\theta)}.$$
Using this together with \eqref{volume_element}, we rewrite \eqref{int_lemm} as
$$I := \int_{\Sigma_0\setminus B_\delta} |x_\xi|^{-p}m(d\xi) = \frac{1}{2\Theta}\int_{\Pi^{-1}(\Sigma_0\setminus B_\delta)}\frac{|x_\xi|^{-p}dx}{|N_\xi|\mu_\xi(\theta)},$$
where $\Pi$ is the mapping defined by \eqref{proj}, and $(\xi,\theta) = \pi^{-1}(x)$.  By \eqref{supp_est}, for appropriate $a > 0$ we have the inclusion $\Pi^{-1}(\Sigma_0\setminus B_\delta)\subset B_{a\delta}^c$. Then, using estimates \eqref{grad_cond}, \eqref{mu_ineq} and the inequality $|x|\leq1$, we obtain
\begin{equation}\label{lemm_internal_estimate}
I\leq C\int_{B^c_{a\delta}}\frac{|x|^{-p}dx}{|N_\xi|\mu_\xi(\theta)}\leq C_1\int_{B^c_{a\delta}}|x|^{-(p+q)}dx\leq C_2 \int_{a\delta}^1 r^{-p-q-1 + d} dr,
\end{equation}
which implies assertion \eqref{int_lemm}.

To prove the second assertion, we again use the inequality \eqref{supp_est} and note that for some  $b > 0$ the inclusion $\Pi^{-1} (\Sigma_0\cap B_\delta)\subset B_{b\delta}$ holds. Now, proceeding with estimates similar to \eqref{lemm_internal_estimate}, we obtain \eqref{int_lemm_2}.

\QEDA

Note that Lemma \ref{lemm_int_over_sigma} implies that the integral
\begin{equation}\label{lemm_corollary}
\int_{\Sigma_0} |x_\xi|^{-p}m(d\xi) < \infty\quad\text{for}\quad p< d-q.
\end{equation}

\begin{lemma}\label{lemm_nbh_property}
Let $0 < \alpha < 1$ be sufficiently small. Then any point $x\in D$ satisfying the inequality
\begin{equation}\label{lemm_ineq}
|x - \Sigma_0|\leq\alpha|x|^q,
\end{equation}
belongs to the set $U_\Theta(\Sigma_0)$.
\end{lemma}
\textit{Proof}. Let us show that for any $x$ satisfying \eqref{lemm_ineq}, the equality
\begin{equation}\label{lemm_eq}
|x-\Sigma_0| = |x - x_\xi|
\end{equation}
holds for some $x_\xi\in\Sigma_0$. By assumption \eqref{D_property} on the set $D$, despite \eqref{abusing_notation}, we have $|x-\Sigma| = |x-a|$ for some $a\in\Sigma$. Then, either $a\in\Sigma_0$ or $a=0$, but the latter is impossible because in that case $|x-\Sigma_0| = |x|$, which contradicts \eqref{lemm_ineq} for $\alpha<1$, due to the inequality $|x|\leq 1$. Thus, the only possibility is $a\in\Sigma_0$, which implies \eqref{lemm_eq}. 

Thus, we have the representation $x = x_\xi+\theta N_\xi$. Then,
\begin{equation}\label{x_repr}
|x - x_\xi| = |\theta N_\xi|
\end{equation}
and
\begin{equation}\label{rem_est}
|\theta N_\xi|\leq \alpha|x|^{q}\leq \alpha|x|.
\end{equation}
Using the triangle inequality, \eqref{x_repr} and \eqref{grad_upper}, we again get \eqref{supp_est}. Combining it with \eqref{rem_est}, we obtain
$$|\theta|\leq C\alpha|x_\xi|^{q}|N_\xi|^{-1} \leq C_2\alpha,$$
where we used \eqref{grad_cond}. Then, choosing $\alpha < \Theta C_2^{-1}$, we get $|\theta| < \Theta$, which proves the lemma. \QEDA

In the following lemma we estimate $|\Omega(x)|$ from below, outside of the neighborhood $U_\Theta(\Sigma_0)$ for small $|x|$.
\begin{lemma}\label{lemm_omega_est}
Let $s>0$ be so small that the ball $B_{2s}\subset D$. Then,
\begin{equation}\label{omega_estimate}
|\Omega(x)|\geq C|x|^{2q},
\end{equation}
uniformly in $x\in B_s\setminus U_\Theta(\Sigma_0)$.
\end{lemma}
\textit{Proof}. For any points $x, x' \in D$ 
\begin{equation}\label{curve_int}
\Omega(x') - \Omega(x) = \int_\gamma(\nabla\Omega(z)\cdot dz),
\end{equation}
where $\gamma\subset D$ is an arbitrary smooth curve connecting points $x$ and $x'$. We take an arbitrary point $x\in B_s\setminus U_\Theta(\Sigma_0)$ and, for definiteness, assume that $\Omega(x) > 0$. Let us consider the curve $\gamma$, which is the unique solution of the equation
\begin{equation}\label{gamma_curve}
\frac{dz}{dt} = -\nabla\Omega(z),\qquad z(0) = x.
\end{equation}
Let $t_0$ be the first time the curve $\gamma=z(t)$ touches the set $\Sigma$ or the boundary of the ball $B_{2s}$
$$t_0 := \min\{t: z(t)\in \Sigma \text{ or } z(t)\in\partial B_{2s}\}.$$
Obviously, $\Omega(z(t_0))\geq0$. Then, setting $x' = z(t_0)$ and using \eqref{curve_int} and \eqref{grad_cond}, we obtain the estimate
\begin{equation}\label{omega_int_est}
\Omega(x)\geq\int_\gamma|\nabla\Omega(z)|dl\geq C\int_\gamma|z|^{q} dl,
\end{equation}
where $dl$ denotes the elementary arc length of the curve $\gamma$. Note that the smallest possible length of the curve $\gamma$ is $|x-\Sigma|$. Indeed, since $0\in\Sigma$, we have $|x-\Sigma|\leq s$, while the distance from $x$ to the boundary of the ball $B_{2s}$ is at least $s$. 

Let us denote by $\gamma_0$ the segment of the curve $\gamma$ length $|x - \Sigma|$, starting at the point $x$. We consider as well a straight-line segment connecting $x$ and $0$ of the same length; denote it by $\Gamma_0$. We choose a natural parameterization on $\gamma_0$ and $\Gamma_0$ such that $\gamma_0(0) = \Gamma_0(0) = x$. Then
\begin{equation}\label{p2p_inequality}
|\gamma_0(\tau)|\geq|\Gamma_0(\tau)|.
\end{equation}
Indeed, since $\Gamma_0$ is a straight-line segment,
$$|\gamma_0(\tau) - x|\leq |\Gamma_0(\tau) - x|,\quad|\Gamma_0(\tau) - x| = |x| - |\Gamma_0(\tau)|.$$
On the other hand,
$$|\gamma_0(\tau) - x|\geq |x| - |\gamma_0(\tau)|,$$
so \eqref{p2p_inequality} follows. We get
$$
\int_{\gamma}|z|^q dl \geq \int_{\gamma_0}|z|^q dl\geq\int_{\Gamma_0}|z|^q dl = \int_{|x| - |x-\Sigma|}^{|x|}r^q dr,
$$
where the last equality uses the fact that $\Gamma_0$ is a radial segment of a ball centered at the origin.
Since $x\in B_s\setminus U_\Theta(\Sigma_0)$, by Lemma \ref{lemm_nbh_property} we have $|x-\Sigma|> \alpha|x|^q$. Consequently,
$$\int_{|x| - |x-\Sigma|}^{|x|}r^q dr > \int_{|x| - \alpha|x|^q}^{|x|}r^q dr\geq C_1|x|^{2q},$$
Together with \eqref{omega_int_est}, this implies \eqref{omega_estimate}. The case $\Omega(x) < 0$ is considered analogously.

\QEDA

\section{Proof of Theorem 1.1}
Let us introduce the parameter
\begin{equation}\label{delta_def}
\delta := \kappa{\nu}^{1/2q},
\end{equation}
where the constant $\kappa = \kappa(\Theta)$, independent of $\nu$, will be chosen later to be sufficiently large. Furthermore, let us introduce the notation
\begin{equation}\label{F_notation}
F_g := F\sqrt{\det g}.
\end{equation}
Obviously, $F_g$ is a $C^2$-smooth function. Let us split the set $D$ into three subsets
\begin{equation}\label{set_separation}
D = \mathfrak{D}_1\cup \mathfrak{D}_2\cup\mathfrak{D}_3,
\end{equation}
where
\begin{equation*}
\mathfrak{D}_1 = B_\delta,\qquad\mathfrak{D}_2 = B^c_\delta\setminus U_\Theta(\Sigma_0),\qquad\mathfrak{D}_3 = B^c_\delta\cap U_\Theta(\Sigma_0).
\end{equation*}
Then the integral $I_\nu$ from \eqref{int_chart} can be represented as
\begin{equation}\label{int_to_estimate}
I_\nu = \sum_{n=1}^3 I_\nu^n,\quad\text{where}\quad I_\nu^n := \int_{\mathfrak{D}_n}\frac{F_g(x)\cos(\rho\nu^{-1}\Omega(x))}{\Omega^2(x) + \nu^2}dx.
\end{equation}
In the sequel we will show that the integrals $I_\nu^1$ and $I_\nu^2$ are bounded above by the quantity $C\chi_{d,4q}(\nu)$, while $I_\nu^3$ yields asymptotic \eqref{int_asymp}.
\subsection{Estimates for the integrals $I_\nu^1$ and $I_\nu^2$}
Since $d\geq 4q$, the integral $I^1_\nu$ can be estimated trivially: 
\begin{equation}\label{int_estimate_1}
|I_\nu^1|\leq C_1\delta^d\nu^{-2} = C\kappa^d\nu^{(d/2q) - 2}\leq C_2.
\end{equation}
To analyze the integral $I^2_\nu$, we introduce a parameter $\beta$ independent of $\nu$ (we assume $\delta < \beta$), and decompose the set  $\mathfrak{D}_2$ as $\mathfrak{D}_2=\hat{\mathfrak{D}}_1\cup\hat{\mathfrak{D}}_2$, where
$$\hat{\mathfrak{D}}_1 = \mathfrak{D}_2\setminus B_\beta,\qquad \hat{\mathfrak{D}}_2 = \mathfrak{D}_2\cap B_\beta.$$
When integrating over the set $\hat{\mathfrak{D}}_1$, the function $\Omega(x)$ is separated from zero, so the integral is bounded by a $\nu$-independent constant. To estimate the integral over the set $\hat{\mathfrak{D}}_2$, we choose sufficiently small $\beta$  and use Lemma~\ref{lemm_omega_est}:
\begin{equation*}
\left|\int_{\hat{\mathfrak{D}}_2}\frac{F_g(x)\cos(\rho\nu^{-1}\Omega(x))dx}{\Omega^2(x) + \nu^2}\right|\leq C\int_{B_\beta\setminus B_\delta}\frac{dx}{|x|^{4q}}\leq C_1 \int_\delta^\beta \frac{r^{d-1}}{r^{4q}}dr\leq C_2\chi_{d, 4q}(\delta).
\end{equation*}
Thus, 
\begin{equation}\label{int_estimate_2}
|I_\nu^2| \leq C\chi_{d, 4q}(\delta) = C_1\chi_{d, 4q}(\nu).
\end{equation}
\subsection{Estimate for the integral $I_\nu^3$}
It suffices to analyze the integral
\begin{equation}\label{I_delta}
I_\delta := \int_{W_\delta}\frac{F_g(x)\cos(\rho\nu^{-1}\Omega(x))dx}{\Omega^2(x) + \nu^2},
\end{equation}
where
\begin{equation}\label{W_set}
W_\delta = \{x = \pi(\xi,\theta)\in U_\Theta(\Sigma_0): x_\xi\in\Sigma_0\setminus B_\delta,\quad|\theta|<\Theta\},
\end{equation}
and $\delta$ is defined in \eqref{delta_def}. Indeed, by inequality \eqref{supp_est}, the inclusion  $W_\delta\,\Delta\,\mathfrak{D}_3\subset B_{c\delta}$ holds for some $c>0$. Then,  
\begin{equation}\label{int_estimate_3}
\left|I_\nu^3 - I_\delta\right| = \left|\int_{W_\delta\,\Delta\,\mathfrak{D}_3}\frac{F_g(x)\cos(\rho\nu^{-1}\Omega(x))dx}{\Omega^2(x) + \nu^2}\right|\leq \left|\int_{B_{c\delta}}\frac{F_g(x)\cos(\rho\nu^{-1}\Omega(x))dx}{\Omega^2(x) + \nu^2}\right|\leq C,
\end{equation}
since the integral over the ball $B_{c\delta}$ can be trivially estimated similarly to \eqref{int_estimate_1}. We rewrite the integral $I_\delta$ in the coordinates $(\xi,\theta)$
\begin{equation}\label{Q_int_xi_theta}
I_\delta = \int_{\Sigma_0\setminus B_\delta} m(d\xi)Y_\nu(\xi),
\end{equation}
where
$$Y_\nu(\xi) := \int_{-\Theta}^\Theta\frac{|N_\xi|\Phi(\xi,\theta)\cos(\rho\nu^{-1}\Omega(\xi,\theta))d\theta}{\Omega^2(\xi,\theta) + \nu^2} = \int_{-\Theta}^\Theta\frac{|N_\xi|\Phi(\xi,\theta)\cos(\rho\nu^{-1}\theta|N_\xi|^2h_\xi(\theta))d\theta}{\theta^2|N_\xi|^4h_\xi^2(\theta) + \nu^2}.$$
Here we used the representation \eqref{omega_repr}, and $\Phi(\xi,\theta) := \mu_\xi(\theta)F_g(\xi,\theta)$. Let us change the integration variable $\theta\rightarrow u$:
$$u = S_\xi(\theta) := \theta h_\xi(\theta),$$
where $h_\xi(\theta)$ is the function from Lemma \ref{unique_param}. By \eqref{g_property}, this change of variables is well defined, provided $\Theta$ is sufficiently small. Furthermore, note that $\sgn S_\xi(\theta) =\sgn\theta$. After the change of variables, the integral takes the form
$$Y_\nu(\xi) = |N_\xi|^{-3}\int_{S_\xi(-\Theta)}^{S_\xi(\Theta)}\frac{\hat {\Phi}(\xi,u)\cos(\rho\eps_\xi^{-1} u)du}{u^2 + \eps_\xi^2},$$
where $\hat{\Phi}(\xi, u) := \Phi(\xi,S_\xi^{-1}(u))/S'_\xi(S^{-1}_\xi(u))$ and $\eps_\xi:=\nu|N_\xi|^{-2}$.
Consider the integral $Y^0_\nu(\xi)$, obtained from $Y_\nu(\xi)$ by replacing the function $\hat\Phi(\xi,u)$ with its value at $u = 0$:
\begin{equation}
Y^0_\nu(\xi) := \frac{\hat \Phi(\xi, 0)}{|N_\xi|^3}\int_{S_\xi(-\Theta)}^{S_\xi(\Theta)}\frac{\cos(\rho\eps_\xi^{-1}u)}{u^2 + \eps_\xi^2} du= \frac{\nu^{-1}\hat\Phi (\xi, 0)}{|N_\xi|}\int_{S_\xi(-\Theta)/\eps_\xi}^{S_\xi(\Theta)/\eps_\xi}\frac{\cos(\rho v)}{v^2 + 1} dv.
\end{equation}
We have
$$\int_{S_\xi(-\Theta)/\eps_\xi}^{S_\xi(\Theta)/\eps_\xi}\frac{\cos(\rho v)}{v^2 + 1} dv = \int_{-\infty}^{\infty}\frac{\cos(\rho v)}{v^2 + 1} dv - \int_{S_\xi(\Theta)/\eps_\xi}^{\infty}\frac{\cos(\rho v)}{v^2 + 1} dv - \int_{-\infty}^{S_\xi(-\Theta)/\eps_\xi}\frac{\cos(\rho v)}{v^2 + 1} dv.$$
The integral in the first term is easily computed:
$$\int_{-\infty}^{\infty}\frac{\cos(\rho v)}{v^2 + 1} dv = \int_{-\infty}^{\infty}\frac{e^{i\rho v}}{v^2 + 1} dv = \pi e^{-|\rho|}.$$
The remaining terms are bounded from above:
$$\left|\int_{-\infty}^{S_\xi(-\Theta)/\eps_\xi}\frac{\cos(\rho v)}{v^2 + 1} dv\right|\leq \int^{S_\xi(-\Theta)/\eps_\xi}_{-\infty}\frac{1}{v^2} dv = \frac{\eps_\xi}{|S_\xi(-\Theta)|}\leq C\eps_\xi,$$
and similarly
$$\left|\int_{S_\xi(\Theta)/\eps_\xi}^{\infty}\frac{\cos(\rho v)}{v^2 + 1} dv\right|\leq C\eps_\xi.$$
Thus, taking into account the relation $\hat\Phi(\xi,0)=F_g(\xi,0)$, we obtain
\begin{equation}\label{Yint_def}
Y_\nu^0(\xi) = \nu^{-1}F_g (\xi, 0)|N_\xi|^{-1}(\pi e^{-|\rho|} + R),\qquad |R|\leq C\nu |N_\xi|^{-2}.
\end{equation}
Next, we estimate the difference between the integrals $Y_\nu^0(\xi)$ and $Y_\nu(\xi)$. We have
$$\hat\Phi(\xi, u) - \hat\Phi(\xi, 0) = A(\xi)u + B(\xi, u)u^2,$$
where the functions $A$ and $B$ are bounded due to the $C^2$-smoothness of the function $\Phi$ and estimates \eqref{g_property}. Then
\begin{equation*}
Y_\nu(\xi) - Y^0_\nu(\xi) = |N_\xi|^{-3}\bigg[\int_{S_\xi(-\Theta)}^{S_\xi(\Theta)}\frac{A(\xi)u\cos(\rho\eps_\xi^{-1} u)du}{u^2 + \eps_\xi^2}
+ \int_{S_\xi(-\Theta)}^{S_\xi(\Theta)}\frac{B(\xi,u)u^2\cos(\rho\eps_\xi^{-1} u)du}{u^2 + \eps_\xi^2}\bigg].
\end{equation*}
For definiteness, assume that $|S_\xi(\Theta)|\geq |S_\xi(-\Theta)|$. Since the integrand in the first term is an odd function and $S_\xi(\Theta) >0$, $S_\xi(-\Theta) < 0$,
$$\left|\int_{S_\xi(-\Theta)}^{S_\xi(\Theta)}\frac{A(\xi)u\cos(\rho\eps_\xi^{-1} u)du}{u^2 + \eps_\xi^2}\right| = \left|\int_{-S_\xi(-\Theta)}^{S_\xi(\Theta)}\frac{A(\xi)u\cos(\rho\eps_\xi^{-1} u)du}{u^2 + \eps_\xi^2}\right|\leq C,$$
where the inequality holds because $0\notin [-S_\xi(-\Theta),S_\xi(\Theta)]$. On the other hand,
$$\left|\int_{S_\xi(-\Theta)}^{S_\xi(\Theta)}\frac{B(\xi,u)u^2\cos(\rho\eps_\xi^{-1} u)du}{u^2 + \eps_\xi^2}\right|\leq C\left|\int_{S_\xi(-\Theta)}^{S_\xi(\Theta)}\frac{u^2}{u^2 + \eps^2_\xi}\right|\leq C_1.$$
Therefore,
\begin{equation}\label{Yint_diff}
|Y_\nu(\xi) - Y^0_\nu(\xi)|\leq C_2 |N_\xi|^{-3}.
\end{equation}
Taking into account estimates \eqref{Yint_def}, \eqref{Yint_diff} and using \eqref{Q_int_xi_theta}, we write
\begin{equation}\label{prepreestimate}
\left| I_\delta - \nu^{-1} e^{-|\rho|}I_\delta^0\right| \leq C\int_{\Sigma_0\setminus B_\delta} m(d\xi)|N_\xi|^{-3} \leq C\chi_{d,4q}(\nu),
\end{equation}
where the last estimate follows from Lemma 2.3 and we introduced the notation
\begin{equation}\label{I_delta_zero}
I_\delta^0 := \pi\int_{\Sigma_0\setminus B_\delta}\frac{F_g(\xi,0) m(d\xi)}{|N_\xi|}.
\end{equation}
Using estimates \eqref{int_estimate_1}, \eqref{int_estimate_2} and \eqref{int_estimate_3}, from \eqref{prepreestimate} we obtain
\begin{equation}\label{preestimate}
\left| I_\nu - \nu^{-1}e^{-|\rho|}I_\delta^0\right|\leq C\chi_{d,4q}(\nu).
\end{equation}
Thus, to conclude the proof, we need to obtain the asymptotics for $I^0_\delta$.
\subsection{Asymptotics for $I_\delta^0$}
Consider the integral
\begin{equation}\label{K_zero}
I_0 = \pi\int_{\Sigma_0}\frac{F_g(\xi,0)m(d\xi)}{|N_\xi|},\qquad |I_0| \leq C\int_{\Sigma_0}|x_\xi|^{-q}m(d\xi)< \infty,
\end{equation}
since the integral on the right-hand side is of the form \eqref{lemm_corollary}. Using \eqref{int_lemm_2}, we get
\begin{equation}\label{K_zero_est}
|I_\delta^0 - I_0| \leq \pi\int_{\Sigma_0\cap B_\delta}\frac{|F_g(\xi,0)|m(d\xi)}{|N_\xi|}\leq C\int_{\Sigma_0\cap B_\delta}|x_\xi|^{-q}m(d\xi)\leq C_1 \delta^{2q}\leq C_2\nu.
\end{equation}
Combining \eqref{preestimate} with \eqref{K_zero_est}, we obtain
\begin{equation}\label{I_a_final}
|I_\nu - \nu^{-1}e^{-|\rho|}I_0|\leq C\chi_{d,4q}(\nu).
\end{equation}
Recalling notation \eqref{F_notation}, we write the integral $I_0$ in the form
$$I_0 = \pi\int_{\Sigma_0}\frac{ F(x_\xi)\sqrt{\det{g(x_\xi)}}m(d\xi)}{|\nabla\Omega(x_\xi)|}.$$
Arguing similarly to the proof of Lemma \ref{unique_param}, it can be shown that in some neighborhood of $\Sigma_0$, there exists a coordinate system such that for any $x_\xi \in\Sigma_0$ and $i\in2,...,d$, the basis vectors $e^i$ lie in $T_\xi\Sigma_0$, while $e^1 \perp_g T_\xi\Sigma_0$, where the subscript $g$ indicates orthogonality with respect to the metric $g$. Without loss of generality, we assume that local coordinates \eqref{loc_coord} are chosen with respect to this basis. This implies that for any $\xi$, the only non-zero element in the first row and first column of $g(x_\xi)$ is the diagonal one, so
$$\det g(x_\xi) = g_{11}(x_\xi)\det g_{\Sigma_0}(x_\xi),$$
where by $g_{\Sigma_0}$ we denote the induced metric on the submanifold $\Sigma_0$. Furthermore, since $\partial_{x^i}\Omega(x_\xi) = 0$ for any $i\geq 2$, on $\Sigma_0$ we have
$$||d\Omega||_g = \sqrt{\sum_{i,j}g^{ij}\partial_{x^i}\Omega\partial_{x^j}\Omega} = \sqrt{g^{11}}|\partial_{x^1}\Omega| = \frac{|\nabla\Omega|}{\sqrt{g_{11}}},$$
where, as before, $g^{ij}$ denotes the elements of the inverse metric tensor. Then, the integral $I_0$ can be rewritten in the form \eqref{int_asymp}, where $\omega_{\Sigma_0} = \sqrt{\det{g_{\Sigma_0}(x_\xi)}}m(d\xi)$ is the volume form on the manifold $\Sigma_0$. Combined with \eqref{I_a_final}, this yields \eqref{statement}.

\QEDA
\section{Appendix}
Let $x,y,z,k\in\mathbb{T}^n$, where $\mathbb{T}^n$ is a $n$-dimensional torus of period $2\pi$, $n\geq 1$. In this section we study the function $\Omega_k$ from \eqref{delta_omega}, corresponding to the case of four-wave interaction, when the dispersion relation $\lambda$ is given by \eqref{omega_kupiainen}
\begin{equation}\label{omega_4wave}
\Omega_k(x,y) := \lambda(x) + \lambda(y) - \lambda(z) - \lambda(k),\quad\text{where}\quad z = x +y - k.
\end{equation}
\begin{lemma}\label{omega_degenerate}
1. For any $k\in\mathbb{T}^n$, coordinates of the critical points of the function $\Omega_k(x, y, z)$ corresponding to the dispersion relation \eqref{omega_kupiainen} satisfy one of the following relations:
\begin{equation}\label{crit_elements}
\begin{split}
&\text{1. }z^i = k^i,\quad x^i = y^i = -k^i,\\
&\text{2. }x^i = -k^i,\quad y^i = \pi + k^i,\quad z^i = \pi - k^i,\\
&\text{3. }x^i = \pi+k^i,\quad y^i = -k^i,\quad z^i = \pi - k^i,\\
&\text{4. }x^i = y^i = \pi - \frac{k^i}{3} - \frac{2\pi j}{3},\quad z^i = -\frac{k^i}3- \frac{2\pi j}{3},\quad j=0,1,2,\
\end{split}
\end{equation}
2. Among the critical points of the function $\Omega_k$ there are degenerate critical points if for some $i$ we have $k^i=\frac{\pi}{2}+\pi j, j\in \mathbb{Z}$. \newline
3. Estimate \eqref{grad_cond_manifold} holds with $q=2$ in some neighborhood of each critical point,uniformly with respect to the choice of the parameter $k$ and the critical point $(x,y)$.
\end{lemma}
\textit{Proof.} For convenience, we make the substitutions $z \to -z$ and $k \to -k$. Then the system of equations for the critical points takes the form
\begin{equation}\label{critical_system}
\partial_{x^i}\Omega = \sin x^i+\sin z^i = 0,\quad
\partial_{y^i}\Omega = \sin y^i + \sin z^i = 0,
\end{equation}
where $i=1,...,n$. From system \eqref{critical_system} it follows that for each $i$,
$$\sin x^i = \sin y^i = -\sin z^i.$$
Thus, for each $i$, up to permutation of $x^i$ and $y^i$, one of the following relations hold on the torus~$\mathbb{T}^n$:
\begin{equation}\label{k_relations}
\begin{split}
&\text{1. } x^i = y^i = -z^i;\qquad\text{2. } x^i = -z^i, \quad y^i = \pi + z^i;\qquad\text{3. } x^i = y^i = \pi+ z^i.\
\end{split}
\end{equation}
Expressing the variables $x^i,y^i,z^i$ in terms of the parameters $k^i$ for each of the relations in \eqref{k_relations}, we obtain \eqref{crit_elements}.

For brevity, we will henceforth omit the subscript $k$ from the notation $\Omega_k$. The Hessian matrix has a block structure, and straightforward calculations show that its determinant is given by
\begin{equation}\label{hessian}
\det\text{Hess}(\Omega(x,y)) = \prod_{i=1}^n\left((\cos x^i - \cos z^i)(\cos y^i - \cos z^i) - \cos^2 z^i\right).
\end{equation}
Substituting any of the equalities from \eqref{crit_elements} into \eqref{hessian}, we see that the determinant vanishes if for some $i$ we have $k^i = \frac{\pi}2 + \pi j$, where $j\in\mathbb{Z}$, and the corresponding coordinate of the critical point satisfies, for example, one of the relations 1-3 in \eqref{crit_elements}. Thus, we obtain the second assertion of the lemma.

Let us consider the gradient $\nabla\Omega(x + \Delta x, y + \Delta y)$ in a neighborhood of a critical point $(x,y)$ of the function $\Omega$, where $|\Delta x^i|, |\Delta y^i| < 1$, $i=1,...,n$. To verify assumption \eqref{grad_cond_manifold}, it suffices to show that
\begin{equation}\label{desired_est}
|\nabla\Omega(x + \Delta x, y + \Delta y)|\geq C(|\Delta x|^2 + |\Delta y|^2)
\end{equation}
for some $C>0$. We fix an index $i$ and write the gradient component corresponding to the $x^i$ direction:
\begin{equation}\label{grad_comp}
\begin{split}
\partial_{x^i}\Omega(x + \Delta x,& y + \Delta y) = \sin x^i \cos\Delta x^i + \cos x^i \sin \Delta x^i\\
&- \sin(x^i + y^i + k^i)\cos(\Delta x^i + \Delta y^i) - \cos(x^i + y^i + k^i)\sin(\Delta x^i + \Delta y^i).
\end{split}
\end{equation}
Using Taylor's formula and the fact that $\sin(x^i) =  \sin(-z^i) = \sin(x^i + y^i + k^i)$, we write
\begin{align}
&\partial_{x^i}\Omega(x + \Delta x, y + \Delta y) = \sin x^i(\cos\Delta x^i - \cos(\Delta x^i + \Delta y^i))\nonumber\\
&+ \cos x^i\sin \Delta x^i - \cos z^i(\sin \Delta y^i \cos\Delta x^i + \sin \Delta x^i \cos\Delta y^i) \nonumber \\
&= \sin x^i\left(\Delta x^i\Delta y^i + \frac{(\Delta y^i)^2}{2}\right)+ (\cos x^i - \cos z^i)\Delta x^i - \cos z^i\Delta y^i+ o\left((|\Delta x^i| + |\Delta y^i|)^2\right).
\label{grad_taylor}
\end{align}
Relations \eqref{crit_elements} imply that $|\cos x^i| = |\cos y^i| = |\cos z^i| =: \sigma$. Let us first consider the case $\sigma \leq \zeta\sqrt{(\Delta x^i)^2 + (\Delta y^i)^2}$ for a sufficiently small constant $\zeta > 0$. 
We have
\begin{align}\label{sum_squares}
(\partial_{x^i}\Omega)^2 + (\partial_{y^i}\Omega)^2 > C&\left((\Delta x^i)^4 + (\Delta y^i)^4 + 8(\Delta x^i)^2(\Delta y^i)^2 + 4\Delta x^i(\Delta y^i)^3 + 4(\Delta x^i)^3\Delta y^i\right)\nonumber\\
& + O\left(\zeta(|\Delta x^i| + |\Delta y^i|)^4\right) + o\left((|\Delta x^i| + |\Delta y^i|)^4\right)\geq C_1\left((\Delta x^i)^2 + (\Delta y^i)^2\right)^2,
\end{align}
where we used the fact that $|\sin x^i| = |\sin y^i|\geq C > 0$. Choosing $\zeta$ small enough, one can see that the last estimate follows from the elementary inequality
$$a^4 + b^4 + 8a^2b^2 + 4ab^3 + 4a^3b \geq \frac{(a^2 + b^2)^2}{3},$$
which holds for any $a,b\in\mathbb{R}$. To verify the latter, due to its homogeneity, it suffices to consider $a$ and $b$ satisfying $a^2 + b^2 = 1$. Let $s:= ab$. Then $(a+b)^2 = 1+2s$, and the left-hand side of the inequality equals $(a+b)^4 + 2a^2b^2 = (1+2s)^2 + 2s^2$. We obtain
$$(1+2s)^2 + 2s^2 = 6s^2 + 4s + 1\geq \frac13.$$

Now we assume that $\sigma > \zeta\sqrt{(\Delta x^i)^2 + (\Delta y^i)^2}$. Then \eqref{grad_taylor} together with a similar relation for $\partial_{y^i}\Omega$ imply
\begin{align}\label{sum_squares2}
(\partial_{x^i}\Omega)^2 + (\partial_{y^i}\Omega)^2 = &\left[(\cos x^i - \cos z^i)^2 + \cos^2 z^i\right](\Delta x^i)^2 + \left[(\cos y^i - \cos z^i)^2 + \cos^2 z^i\right](\Delta y^i)^2\nonumber\\
&-2\cos z^i\left[\cos x^i + \cos y^i - 2\cos z^i\right]\Delta x^i\Delta y^i + o\left((|\Delta x^i| + |\Delta y^i|)^2\right)\nonumber\\
&\geq C\cos^2 z^i\left((\Delta x^i)^2 + (\Delta y^i)^2\right)\geq C_1\zeta^2\left((\Delta x^i)^2 + (\Delta y^i)^2\right)^2 \,,
\end{align}
where the first inequality holds due to the positive definiteness of the quadratic form in $\Delta x^i$, $\Delta y^i$ for any $x^i$, $y^i$, $z^i$ satisfying any of the equalities in \eqref{k_relations}. Thus, estimate \eqref{sum_squares} holds for every $i=1,...,n$ uniformly with respect to the choice of $x$, $y$, $z$ and the parameter $k$, which implies \eqref{desired_est}.

\QEDA

%где первое равенство выполнено в силу второго уравнения в \eqref{critical_system}, а последние два %неравенства выполнены в силу \eqref{crit_elements} при $k^i=\pm\frac14$. Таким образом, в окрестности %вырожденных особых точек выполнено $\nabla\Omega^i_1(k_1^i + \Delta k_1^i, k_2^i + \Delta k_2^i) \geq C((\Delta k_2^i)^2 + \Delta k_1^i\Delta k_2^i)$, откуда следует
%$$|\nabla\Omega(k_1 + \Delta k_1, k_2 + \Delta k_2)|\geq C((\Delta k_1)^2 + (\Delta k_2)^2),$$
%что и означает выполнение оценки \eqref{grad_cond} с $q=2$.

\textbf{Acknowledgment.} This work was supported by the Russian Science Foundation under grant no.25-11-00114.

The author is grateful to Andrey Dymov for fruitful discussions.

\end{document}